\documentclass[conference]{IEEEtran}
\IEEEoverridecommandlockouts
\usepackage{cite}
\usepackage{amsmath,amssymb,amsfonts}
\usepackage{algorithmic}
\usepackage{graphicx}
\usepackage{textcomp}
\usepackage{xcolor}
\usepackage{physics}
\usepackage{tikz}
\usepackage{url}
\usepackage{hyperref} 
\def\BibTeX{{\rm B\kern-.05em{\sc i\kern-.025em b}\kern-.08em
    T\kern-.1667em\lower.7ex\hbox{E}\kern-.125emX}}
    
\begin{document}

\title{Benchmarking Quantum and Classical Algorithms for the 1D Burgers' Equation: QTN, HSE, and PINN}

\author{
\IEEEauthorblockN{Vanshaj Kerni}
\IEEEauthorblockA{\textit{Department of Physics} \\
\textit{Indian Institute of Technology Roorkee}\\
Roorkee, India \\
kernivanshaj@gmail.com}
\url{orcid.org/0000-0002-3076-607X}
\and
\IEEEauthorblockN{Abdelrahman E. Ahmed}
\IEEEauthorblockA{\textit{M.Sc. Student, Computer Science} \\
\textit{Alexandria University}\\
Alexandria, Egypt \\
abdo.elsayd102@gmail.com \\
\url{orcid.org/0009-0009-1618-2404}} \\
\and
\IEEEauthorblockN{Syed Ali Asghar}
\IEEEauthorblockA{\textit{Department of Physics} \\
\textit{University of Karachi}\\
Karachi, Pakistan \\
syedaliazgher2001@gmail.com}
\url{orcid.org/0009-0006-5559-6050}
}
\maketitle

\begin{abstract}
We present a comparative benchmark of Quantum Tensor Networks (QTN), the Hydrodynamic Schrödinger Equation (HSE), and Physics-Informed Neural Networks (PINN) for simulating the 1D Burgers' equation. Evaluating these emerging paradigms against classical GMRES and Spectral baselines, we analyse solution accuracy, runtime scaling, and resource overhead across grid resolutions ranging from $N=4$ to $N=128$. Our results reveal a distinct performance hierarchy. The QTN solver achieves superior precision ($L_2 \sim 10^{-7}$) with remarkable near-constant runtime scaling, effectively leveraging entanglement compression to capture shock fronts. In contrast, while the Finite-Difference HSE implementation remains robust, the Spectral HSE method suffers catastrophic numerical instability at high resolutions, diverging significantly at $N=128$. PINNs demonstrate flexibility as mesh-free solvers but stall at lower accuracy tiers ($L_2 \sim 10^{-1}$), limited by spectral bias compared to grid-based methods. Ultimately, while quantum methods offer novel representational advantages for low-resolution fluid dynamics, this study confirms they currently yield no computational advantage over classical solvers without fault tolerance or significant algorithmic breakthroughs in handling non-linear feedback
\end{abstract}


\section{Introduction}
Computational Fluid Dynamics (CFD) has long relied on grid-based techniques such as the Finite Difference Method (FDM) and Spectral methods to solve nonlinear Partial Differential Equations (PDEs) like the Navier-Stokes equations \cite{Anderson1995, Kim1985}. While effective for smooth flows, these classical approaches scale unfavourably with system size $N$, typically becoming computationally prohibitive for high-resolution 3D turbulence simulations. The 1D Burgers' equation \cite{Bateman1915}, which retains the nonlinear convection-diffusion structure of the Navier-Stokes equation, serves as a critical benchmark for evaluating new solvers capable of handling shocks and multi-scale discontinuities.

In recent years, Physics-Informed Neural Networks (PINNs) \cite{Raissi2019} have emerged as a mesh-free alternative, embedding physical laws directly into the optimisation landscape. While PINNs offer flexibility, standard formulations often struggle with the \textit{spectral bias} problem and optimisation difficulties in high-frequency or chaotic regimes \cite{Hu2024}. This limits their scalability for stiff problems compared to high-precision numerical solvers.

Quantum computing offers a fundamentally different approach, potentially overcoming classical bottlenecks through high-dimensional Hilbert space representations. Quantum algorithms like HHL \cite{Harrow2009} and QSVT \cite{Gilyen2019} promise exponential speedups for linear systems, but their application to nonlinear PDEs is non-trivial. Moreover, the input/output bottleneck and the requirement for fault tolerance often negate theoretical gains \cite{Preskill2018, Costa2022, Linden2022}.

To address nonlinearity in the Noisy Intermediate-Scale Quantum (NISQ) era, two distinct paradigms have emerged: Quantum Tensor Networks (QTN) and quantum-native Hamiltonian simulation. QTNs, originally from condensed matter physics \cite{White1992}, operate on classical hardware but use quantum-inspired analytical tensor factorisations to compress the solution space \cite{Lubasch2020}. Conversely, the Hydrodynamic Schr\"odinger Equation (HSE) \cite{Meng2023} utilises the Madelung transformation to map fluid variables to a wavefunction, enabling direct unitary evolution on quantum processors.

In this work, we present a rigorous comparative benchmark of these methods against classical GMRES and Spectral solvers. We explicitly distinguish between quantum-inspired (QTN) and quantum-native (HSE) approaches. Our goal is not to claim quantum advantage, but to diagnose the algorithmic maturity and specific bottlenecks, such as entanglement growth and readout costs, of these emerging solvers. We benchmark on the 1D Burgers' equation, focusing on metrics relevant to the quantum community: circuit depth, entanglement scaling, and noise sensitivity.

The remainder of this work is organised as follows: Section \ref{sec:formulation} details the mathematical formalism and algorithmic foundations; Section \ref{sec:implementation} describes the Setup of the algorithms. Section V reports the comparative results and error analysis, and Section VI concludes with future directions.

\section{Problem Formulation and Algorithmic Background}
\label{sec:formulation}
\subsection{Importance}
The 1D viscous Burgers equation is a canonical nonlinear advection--diffusion model widely used to study wave steepening, shock formation, and dissipative transport mechanisms. The governing equation is
\begin{equation}
\frac{\partial u(x,t)}{\partial t} + u(x,t)\frac{\partial u(x,t)}{\partial x}
= \nu \frac{\partial^{2}u(x,t)}{\partial x^{2}},
\label{eq:burgers}
\end{equation}
where $u(x,t)$ denotes the velocity field and $\nu$ is the kinematic viscosity. The nonlinear convective term $u\,u_x$ drives the steepening of spatial gradients, leading to the emergence of shock-like structure. In contrast, the inviscid regime, whereas the diffusive term $\nu u_{xx}$ provides viscous smoothing. In the limit $\nu \rightarrow 0$, burger equation \eqref{eq:burgers} reduces to a hyperbolic conservation law supporting discontinuous shocks governed by Rankine-Hugoniot conditions. Due to its analytical tractability and structural similarity to the Navier-Stokes equations, the Burgers equation remains a central benchmark for numerical solvers, reduced-order modelling, and modern PDE analysis frameworks \cite{burgers1948,zhang2020}.

\subsection{Formalism}
We consider the viscous 1D Burgers' equation on $x \in [0,1]$ over time $t \in [0, T]$:
\begin{equation}
\frac{\partial u}{\partial t} + u\frac{\partial u}{\partial x}
= \nu \frac{\partial^{2}u}{\partial x^{2}},
\label{eq:burgers_formal}
\end{equation}
where $\nu$ is the kinematic viscosity. We define the Reynolds number as $Re = 1/(2\nu)$. The problem is initialised with a Riemann step profile to induce shock formation:
\begin{equation}
u(x,0)=
\begin{cases}
u_L & 0 \leq x \leq 0.5, \\
u_R & 0.5 < x \leq 1,
\end{cases}
\label{eq:step_initial}
\end{equation}
with Dirichlet boundaries $u(0,t)=\alpha, u(1,t)=\beta$. In our benchmarks, we set $u_L=1.0, u_R=0.0$, $\alpha=1.0, \beta=0.0$, and vary $\nu$ to study stability.

\begin{figure}[htp!]
\label{grid}
\centering 
\begin{tikzpicture}[scale=0.2] 
\definecolor{lightgray}{gray}{0.85} \definecolor{gridgray}{gray}{0.6} %
\def\Nx{32} 
\def\Nt{32}
\foreach \j in {0,...,\numexpr\Nt-1}{ 
    \foreach \i in {0,...,\numexpr\Nx-1}{ 
        \fill[lightgray] (\i,\j) rectangle (\i+1,\j+1); 
    }} 
\foreach \i in {0,...,\Nx} {
    \draw[gray!70] (\i,0) -- (\i,\Nt);
}
\foreach \j in {0,...,\Nt} {
    \draw[gray!70] (0,\j) -- (\Nx,\j);
}
\node[below] at (0,-0.8) {\small $x=0$};
\node[below] at (\Nx,-0.8) {\small $x=L$};
\node[below] at (\Nx/2,-0.8) {\small Spatial grid ($N$ points)};
\node[left] at (-0.8,0) {\small $t=0$};
\node[left] at (-0.8,\Nt) {\small $t=T$};
\node[rotate=90] at (-2,\Nt/2) {\small Time evolution ($T$ steps)};
\node[above] at (0, \Nt+0.3) {$u=\alpha$ (Dirichlet)};
\node[above] at (\Nx, \Nt+0.3) {$u=\beta$ (Dirichlet)};
\node at (\Nx/2, \Nt+\Nt/8) {\small Space--time discretised for 1D Burgers' equation};
\end{tikzpicture}
\end{figure}

To ensure a fair comparison, all grid-based methods (GMRES, Spectral, QTN, HSE) use the same spatial resolution $N$ (where $N=2^n$ for quantum methods) and identical time integration horizons based on the Courant-Friedrichs-Lewy (CFL) condition \cite{Courant1928-CFL}, defined as
\begin{equation}
    \Delta t = C_{\text{CFL}} \cdot \min\left( \frac{\Delta x}{\max|u|}, \frac{\Delta x^2}{\nu} \right); \ {C_{\text{CFL}} = 0.1}
\end{equation}
where $\nu$ is the viscosity, and $\max|u|$ is the peak velocity magnitude in the current state. Error is quantified using the relative $L_2$ norm:
\begin{equation}
\mathcal{E} = \frac{\|u_{\text{pred}} - u_{\text{ref}}\|_2}{\|u_{\text{ref}}\|_2},
\end{equation}
computed at the final time $T$. For HSE, we account for the readout cost of reconstructing $u(x,t)$ from the wavefunction requires $\mathcal{O}(N)$ measurements. 

Five numerical discretisation strategies were evaluated, over a spatial grid, visualised in Figure \ref{grid}, ranging from coarse to fine resolutions with three initial profiles: sine, Gaussian, and smooth step profiles. We computed the discrete $\mathrm{L}_2$ error norm, compared in table \ref{tab:all_discretizations} to calculate the numerical accuracy of different methods. The discretisation strategy was selected based on its achieved $\mathrm{L}_2$ error performance and practical feasibility for implementation within the Pennylane framework.
\begin{table}[ht]
\centering
\caption{L$_2$ error comparison across numerical discretisation methods for different initial conditions.}
\label{tab:all_discretizations}
\begin{tabular}{lll}
\hline
\textbf{Method} & \textbf{Test Case} & \textbf{L$_2$ Error} \\
\hline
Chebyshev Collocation   & Sine Wave      & $1.15 \times 10^{-2}$ \\
                        & Gaussian Pulse & $1.64 \times 10^{-1}$ \\
\hline
Discontinuous Galerkin  & Sine Wave      & $8.36 \times 10^{-2}$ \\
                            & Gaussian Pulse & $1.39 \times 10^{-1}$ \\
                            & Smooth Step    & $2.43 \times 10^{-1}$ \\
\hline
Finite Difference Method & Sine Wave      & $6.50 \times 10^{-2}$ \\
                         & Gaussian Pulse & $1.39 \times 10^{-1}$ \\
                         & Smooth Step    & $2.78 \times 10^{-1}$ \\
\hline
Finite Element Method    & Sine Wave      & $1.08 \times 10^{-2}$ \\
                         & Gaussian Pulse & $1.24 \times 10^{-1}$ \\
                         & Smooth Step    & $2.76 \times 10^{-1}$ \\
\hline
Finite Volume Method      & Sine Wave      & $4.79 \times 10^{-2}$ \\
                         & Gaussian Pulse & $1.15 \times 10^{-1}$ \\
                         & Smooth Step    & $2.59 \times 10^{-1}$ \\
\hline
Fourier Spectral         & Sine Wave      & $4.69 \times 10^{-3}$ \\
                         & Gaussian Pulse & $9.46 \times 10^{-2}$ 
                         \\
\hline
\end{tabular}
\end{table}
For the baseline, we implemented a GMRES solver, providing a high-fidelity reference solution for error assessment and convergence analysis. We also implemented a PINN-based approach, comparing a neural network-based approach, adding another benchmark to quantum algorithms. 

\section{Setup}
\label{sec:implementation}
\subsection{Classical GMRES solver}
The classical GMRES solver acted as the baseline for the study. The discretisation method for the nonlinear convection term, $uu_x$, is identical to other approaches for consistency. The resulting sparse linear system, $\big(I - \nu\,\Delta t\,\mathcal{L}\big)\,u^{n+1} = u^{n} - \Delta t\,\mathcal{C}(u^{n})$, is solved using the Generalised Minimal RESidual method (GMRES) \cite{2020SciPy-NMeth}, with boundary conditions implemented via SciPy \cite{Virtanen2020}. To maintain uniformity across grid resolutions, the solver is applied consistently for all spatial sizes considered in this study, ranging from $N = 4$ to $N = 128$.

\subsection{Neural Network Solver -- PINN}
We adopt a standard Physics-Informed Neural Network (PINN) formulation \cite{Raissi2019}. The network approximates $u(x,t)$ using a fully connected architecture with 3 hidden layers of 50 neurons each and hyperbolic tangent ($\tanh$) activation. The loss function $\mathcal{L}_{total} = \mathcal{L}_{PDE} + \mathcal{L}_{IC} + \lambda_{BC}\mathcal{L}_{BC}$ minimises the PDE residual, initial, and boundary discrepancies. Training was performed using the Adam optimizer with a learning rate of \(10^{-3}\), applied across spatial–temporal grids ranging from \(N_x = 50\) to \(N_x = 200\) and \(N_t = 50\) to \(N_t = 200\), and viscosity settings corresponding to \(R \in [50, 100]\) for 10,000 epochs.

\begin{figure}[h]
    \centering
    \includegraphics[width=\columnwidth]{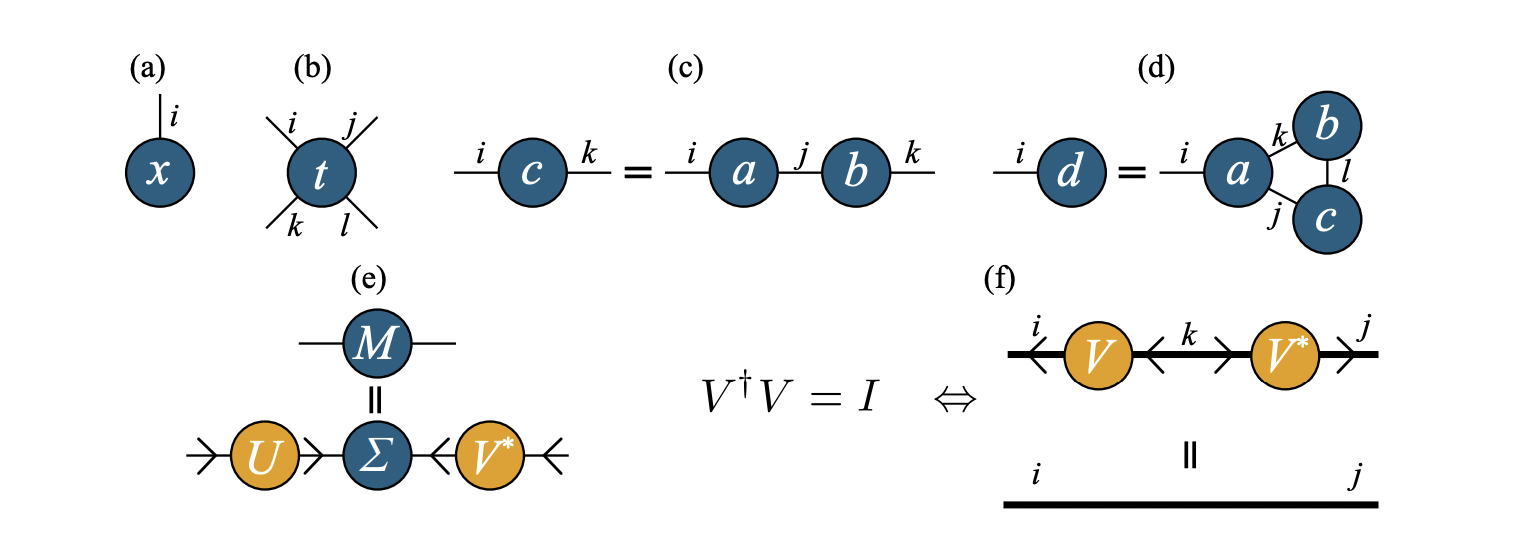}
    \caption{Graphical representations of (a) a vector, (b) a fourth-order tensor, (c) matrix multiplication, (d) the expression in Equation~(2), (e) the singular value decomposition (SVD) of a tensor, and (f) the definition of an isometry. Adapted from the original source, fig. 1.}
    \label{fig:tensor_network}
\end{figure}

\subsection{Quantum Tensor Networks}
\label{qtn_setup}
We used a custom Python-based Quantum Tensor Network (QTN) solver that leverages the Matrix Product State (MPS). Our method directly evolves the quantum state vector as an MPS, offering a resource-efficient representation. The bond dimension $\chi$ governs the amount of entanglement the state can capture. Linear operations are implemented as analytic Matrix Product Operators (MPOs). Non-linear advection term ($u \cdot u_x$) is computed using a site-wise Hadamard product of the state MPS and the derivative MPS. As this operation temporarily increases the bond dimension to $\chi^2$, it is immediately followed by Singular Value Decomposition (SVD) based compression to truncate the state back to a tractable rank.
Time integration is performed using an explicit fourth-order Runge-Kutta (RK4) method with adaptive time-stepping determined by advective and diffusive Courant-Friedrichs-Lewy (CFL) stability constraints. To ensure numerical stability, the solver applies adaptive truncation after each update, discarding singular values below a threshold $\epsilon_{\text{cutoff}}$ while monitoring mass, momentum, and energy conservation.

\subsection{Hydrodynamic Schr\"odinger Equation (HSE)}
The HSE solver utilises the Madelung transformation to map the nonlinear viscous Burgers' equation to a linear Schrödinger-like equation. The velocity field is encoded into an $n$-qubit wavefunction $|\psi\rangle$, evolved under an effective Hamiltonian $\hat{H} = \hat{H}_k + \hat{H}_q$, where, $\hat{H}_k$ represents the kinetic diffusion term, while $\hat{H}_q$ is a density-dependent quantum potential. We considered a spinless fluid element of unit mass ($\tilde m=1$) represented by $\psi(x, t)$ with probability density $\rho = \bar\psi \psi$. Two Hamiltonian discretisation strategies are evaluated: a Finite Difference (FD) approach, yielding a sparse tridiagonal matrix with polynomial gate scaling, and a Fourier-based Spectral method, with superior spatial accuracy resulting in a dense Hamiltonian with exponential circuit depth. The Hamiltonian is decomposed into Pauli strings, $\hat{H} = \sum c_j \hat{P}_j$, and time evolution is implemented via a first-order Trotter-Suzuki decomposition. To assess feasibility in the NISQ era, simulations are conducted using density matrix formalism incorporating depolarising and amplitude damping noise channels. The classical velocity profile is reconstructed at each time step by extracting the phase gradient of the output wavefunction. 

\subsection{Computational Setup}
The numerical simulations were performed on two distinct computational configurations to evaluate performance across different hardware architectures:

\begin{enumerate}
    \item \textbf{High-Performance Workstation:} Intel Core i9 CPU and 64GB of RAM. GPU acceleration for Physics-Informed Neural Network (PINN) training was provided by an NVIDIA RTX 3090. The software environment included Python 3.10, utilising PyTorch 2.0 for PINN models, PennyLane 0.33 for quantum circuit simulations, and SciPy 1.10 for classical solvers.
    
    \item \textbf{Apple Silicon:} Apple M4 chip (10 CPU, 8 GPU cores) with 16GB of unified memory, Hardware-accelerated ray tracing with a 16-core Neural Engine and a 256GB SSD. The software stack was updated to Python 3.12, incorporating PennyLane 0.43.2, SciPy 1.16.3, and PyTorch 2.8.0.
\end{enumerate}

\section{State Implementation \& Evolution}

\subsection{QTN circuit}
\subsubsection*{Implementation}
We mapped the initial vector $\mathbf{u}_0 \in \mathbb{R}^N$ to a Matrix Product State through a sequential Singular Value Decomposition (SVD) algorithm. For a system of size $N=2^L$, the vector is decomposed into a train of $L$ rank-3 tensors $A^{(k)}$, connected by bond indices $\alpha_k$. 
The input vector $\mathbf{u}_0$ is treated as a tensor $C^{(0)}$ of shape $(1, N)$, factorised for each site $k$ from $0$ to $L-2$, isolating the physical index $i_k$. The tensor $C^{(k)}$ is reshaped into a local matrix $M$ of dimensions $(r_k \cdot 2) \times 2^{L-k-1}$. The SVD algorithm is implemented such that $M$ is decomposed into $M = U \Sigma V^\dagger$. To bound the entanglement entropy, we retain only the $\chi_k = \min(\chi_{\max}, \text{rank}(M))$ largest singular values. 

\begin{figure}[htp!]
    \centering
    \includegraphics[width=\columnwidth]{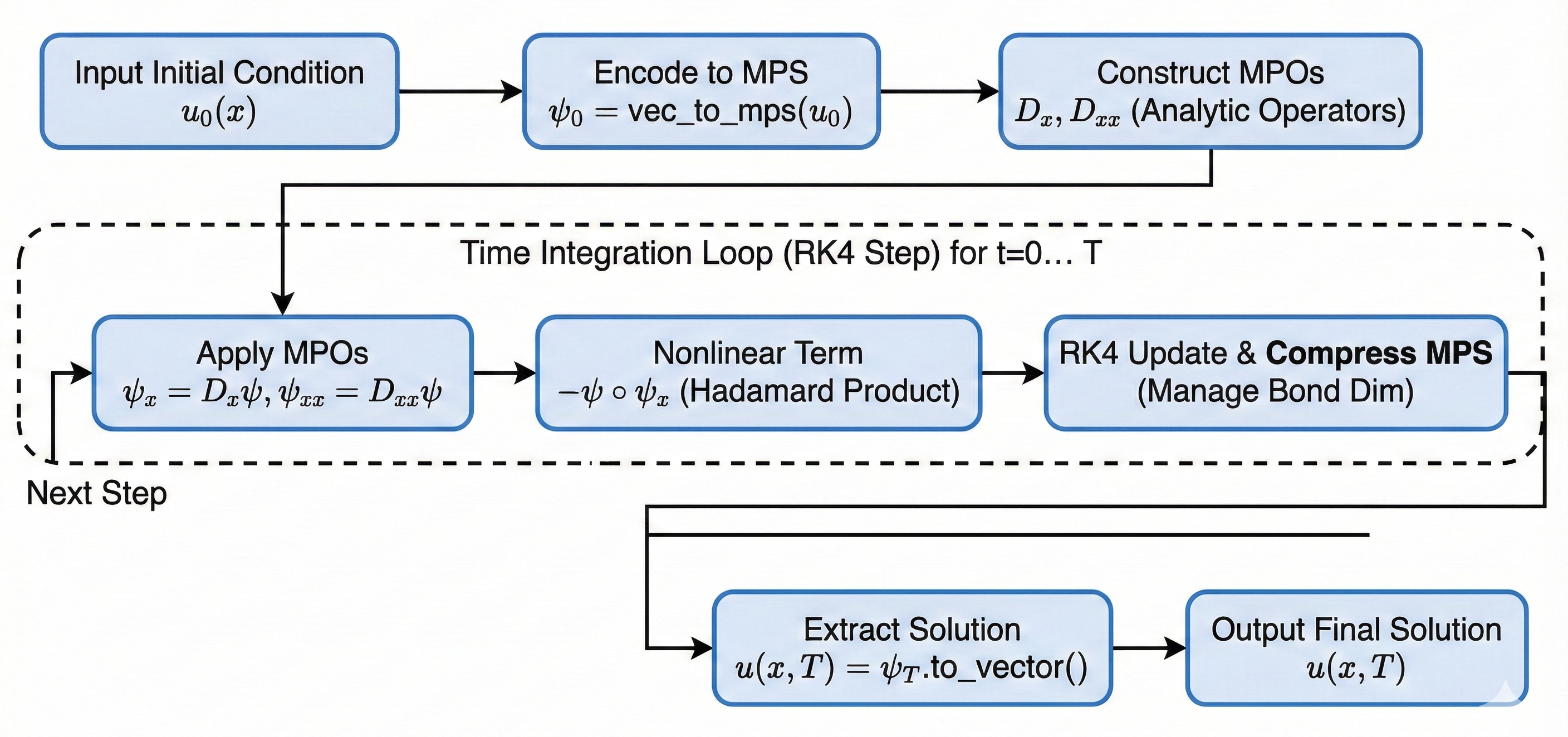}
    \caption{Caption}
    \label{fig:placeholder}
\end{figure}

The unitary matrix $U$ is reshaped to form the MPS tensor $A^{(k)}$ of shape $(r_k, 2, \chi_k)$. The remainder, $\Sigma V^\dagger$, becomes the input $C^{(k+1)}$ for the next site. The final residual $C^{(L-1)}$ is reshaped directly into the last tensor $A^{(L-1)}$ of shape $(r_{L-1}, 2, 1)$.

\subsubsection*{Time Evolution}
Time integration is performed using an explicit fourth-order Runge-Kutta (RK4) scheme. To ensure numerical stability, the time step $\Delta t$ is adapted dynamically at each iteration according to the Courant-Friedrichs-Lewy (CFL) condition \cite{Courant1928-CFL}:
\begin{equation}
    \Delta t = C_{\text{CFL}} \cdot \min\left( \frac{\Delta x}{\max|u|}, \frac{\Delta x^2}{\nu} \right), \quad \text{with } C_{\text{CFL}} = 0.1,
\end{equation}
where $\nu$ is the viscosity and $\max|u|$ represents the peak velocity magnitude in the current state.

A critical component of the algorithm is the evaluation of the nonlinear advection term, which necessitates the Hadamard product of the solution state $|\psi\rangle$ and its gradient $|\psi_x\rangle$. We construct a target MPS representing the product state $|\psi_{\text{nonlin}}\rangle = |\psi\rangle \odot |\psi_x\rangle$. In our implementation, the routine \texttt{hadamard\_product\_controlled} computes the element-wise product of two generic MPS, $|\psi_1\rangle$ and $|\psi_2\rangle$. For each lattice site $j \in \{0, \dots, L-1\}$, the local tensors $A^{(j)} \in \mathbb{R}^{\chi_A \times 2 \times \chi_A'}$ and $B^{(j)} \in \mathbb{R}^{\chi_B \times 2 \times \chi_B'}$ are retrieved. Subsequently, for each physical index $i \in \{0, 1\}$, we compute the Kronecker product of the corresponding bond matrices to generate $M^{(j)}_i$ with dimensions $(\chi_A \chi_B) \times (\chi_A' \chi_B')$:
\begin{equation}
    M^{(j)}_i = A^{(j)}[i] \otimes B^{(j)}[i].
\end{equation}
The new local tensor $T^{(j)}$ is assembled by stacking $M^{(j)}_0$ and $M^{(j)}_1$ along the physical axis, resulting in a tensor of shape $(\chi_A \chi_B) \times 2 \times (\chi_A' \chi_B')$. Finally, SVD truncation is applied to compress the bond dimension back to the target rank $\chi_{\max}$, thereby preventing the exponential growth of entanglement entropy.

\subsection{HSE circuit}
\label{hse_setup}
\subsubsection*{Implementation}
the classical velocity field, $u(x,t=0)$, is mapped to the initial quantum state $|\psi\rangle$ using the inverse Madelung transformation defined in Eq. \ref{eq:inv_madelung}:
\begin{equation}
\psi(x,t) = \sqrt{\rho(x,t)} \exp\left(\frac{i}{\nu} \int u(x,t) \ dx\right)
\label{eq:inv_madelung}
\end{equation}
This state is realised on the PennyLane backend using the \texttt{StatePrep} encoding routine, which encodes the complex amplitudes directly into the computational basis states of the qubit register.

\begin{figure}[htp!]
    \centering
    \includegraphics[width=\columnwidth]{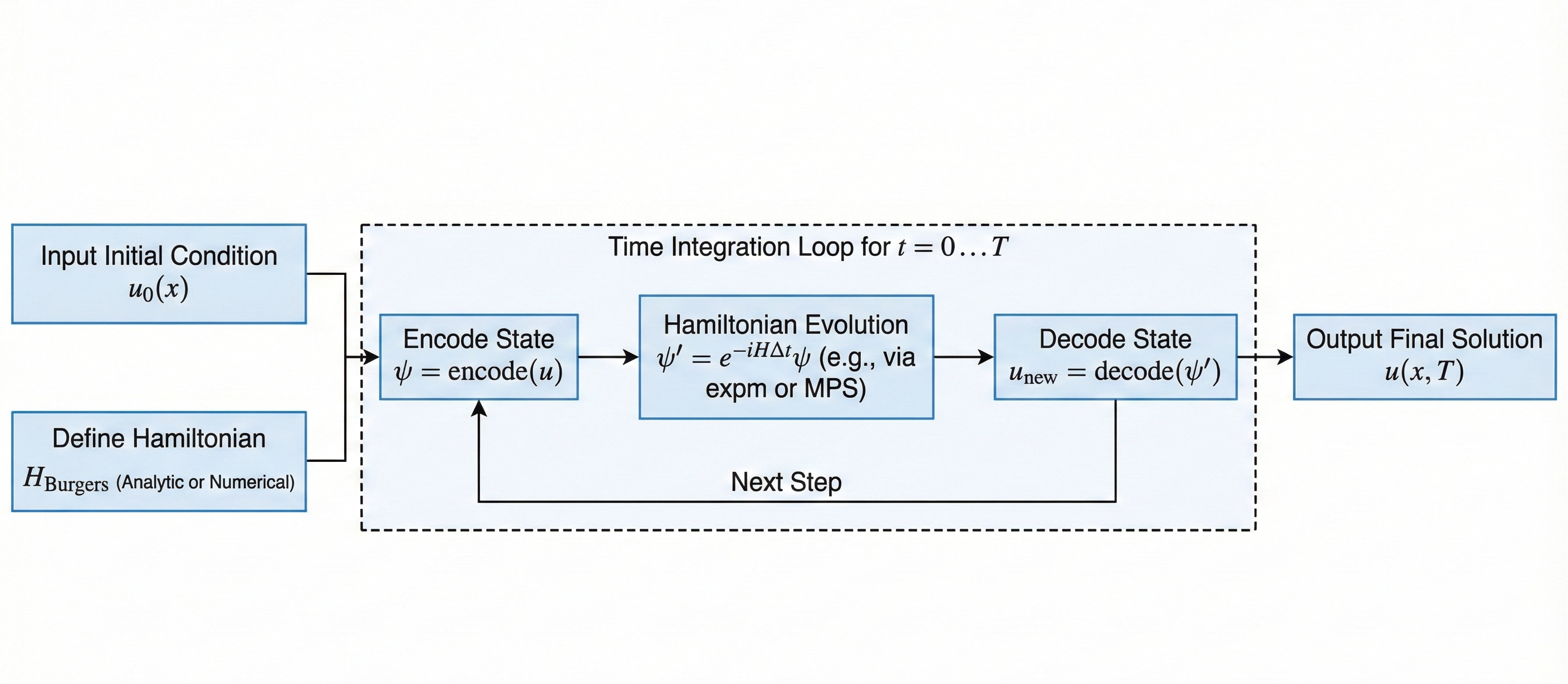}
    \caption{architecture: Schematic architecture of the HSE implementation and time evolution of the state. }
    \label{fig:hse}
\end{figure}

The total Hamiltonian, $H$, is constructed as a Hermitian matrix of dimensions $2^n \times 2^n$. To implement this on gate-based hardware, we decompose $H$ into a weighted sum of Pauli operators:\begin{align}H = \sum_{j=1}^M c_j P_j\end{align}where $c_j \in \mathbb{R}$ are the coefficients and $P_j$ are $n$-qubit Pauli strings ($P_j \in \{I, X, Y, Z\}^{\otimes n}$). This decomposition is performed classically for both the Finite Difference and Spectral Hamiltonian constructions using the \texttt{pauli\_decompose} routine, yielding a \texttt{qml.Hamiltonian} object.

\subsubsection*{Time Evolution}
The time evolution, $U(t) = e^{-iHt}$, is implemented by two distinct evolution strategies, a first-order Trotter-Suzuki product formula and a variational Parameterised Trotter approach. The first-order Trotter-Suzuki evolution is employed by a fixed small time step $c_j \Delta t$, approximated by sequential exponentials of the non-commuting kinetic ($\hat{T}$) and potential ($\hat{Q}$) terms. In PennyLane, each exponential factor is implemented as a \texttt{qml.Exp} gate. This sequence is repeated for $m$ steps to simulate the full trajectory from $t=0$ to $t=T$.
\begin{equation}
e^{-i(\hat{T} + \hat{Q})\Delta t} \approx \prod_{j=1}^M e^{-i c_j P_j \Delta t}
\end{equation}
The Parameterised Trotter replaces the fixed rotation angles $\phi_j = c_j \Delta t$ with a set of trainable parameters $\boldsymbol{\theta}$. The new unitary operator, $U(\boldsymbol{\theta})$, is implemented via the \texttt{qml.PauliRot} gate\footnote{the gate argument is scaled by a factor of 2 to account for the standard rotation definition $R_P(\phi) = e^{-i \phi P / 2}$}
Consisting of L layers, each layer is applied with a product of parameterised Pauli rotations, corresponding to the terms in the Hamiltonian decomposition. 
\begin{equation} U(\boldsymbol{\theta}) = \prod_{l=1}^{L} \prod_{j=1}^{M} \exp\left(-i \theta_{l,j} P_j\right) 
\end{equation}
where $\theta_{l,j}$ is the trainable rotation angle for the $j$-th Pauli term in the $l$-th layer.

We utilise the Adam optimiser with a learning rate of $\eta = 0.05$ to minimise the $L_2$ loss between variational state $|\psi(\boldsymbol{\theta})\rangle = U(\boldsymbol{\theta})|\psi_0\rangle$ and a high-precision reference target $|\psi_{\text{target}}\rangle = e^{-iH\Delta t}|\psi_0\rangle$ with the loss function defined, 
\begin{equation}
    \mathcal{L}(\boldsymbol{\theta}) = \big\| |\psi(\boldsymbol{\theta})\rangle - |\psi_{\text{target}}\rangle \big\|^2 
\end{equation}
To accelerate convergence, parameters are initialised using the standard Trotter angles plus a small perturbation: $\theta_{l,j}^{(0)} = c_j \Delta t + \epsilon$. This allows for compressing the simulation error into a shallower circuit depth compared to the standard decomposition. 

\begin{table}[htbp]
\caption{Comparison of Solvers ($N$: grid size, $n = \log_2 N$)}
\begin{center}
\begin{tabular}{|c|c|c|c|c|}
\hline
\textbf{Method} & \textbf{Error} & \textbf{Runtime} & \textbf{Memory} & \textbf{Qubits} \\
\hline
GMRES & $\sim 10^{-4}$ & $\mathcal{O}(N)$ & $\mathcal{O}(N)$ & N/A \\
QTN & $\sim 10^{-4}$ & $\mathcal{O}($n$ \cdot \chi^4)$ & $\mathcal{O}($n$ \cdot \chi^2)$ & $n$ \\
HSE & $\sim 10^{-3}$ & $\mathcal{O}(poly(n) \cdot 1/\epsilon)$ & $\mathcal{O}(2^n)$ & $n$ \\
PINN & $\sim 10^{-2}$ & Training dependent & $\mathcal{O}(weights)$ & N/A \\
\hline
\end{tabular}
\label{runtime_complexity}
\end{center}
\end{table}

\section{Results and Discussion}
\label{sec:results}
We evaluate the quantum tensor network \ref{qtn_setup} and the hydrodynamic schr\"odinger equation \ref{hse_setup}, against classical baselines. The benchmarking framework utilises the 1D Burgers' equation with viscosity $\nu=0.01$, evolving the system over a total time $T=1.0$ with a time step $dt=0.005$. 

\begin{figure}[ht!]
    \centering
    \includegraphics[width=\columnwidth]{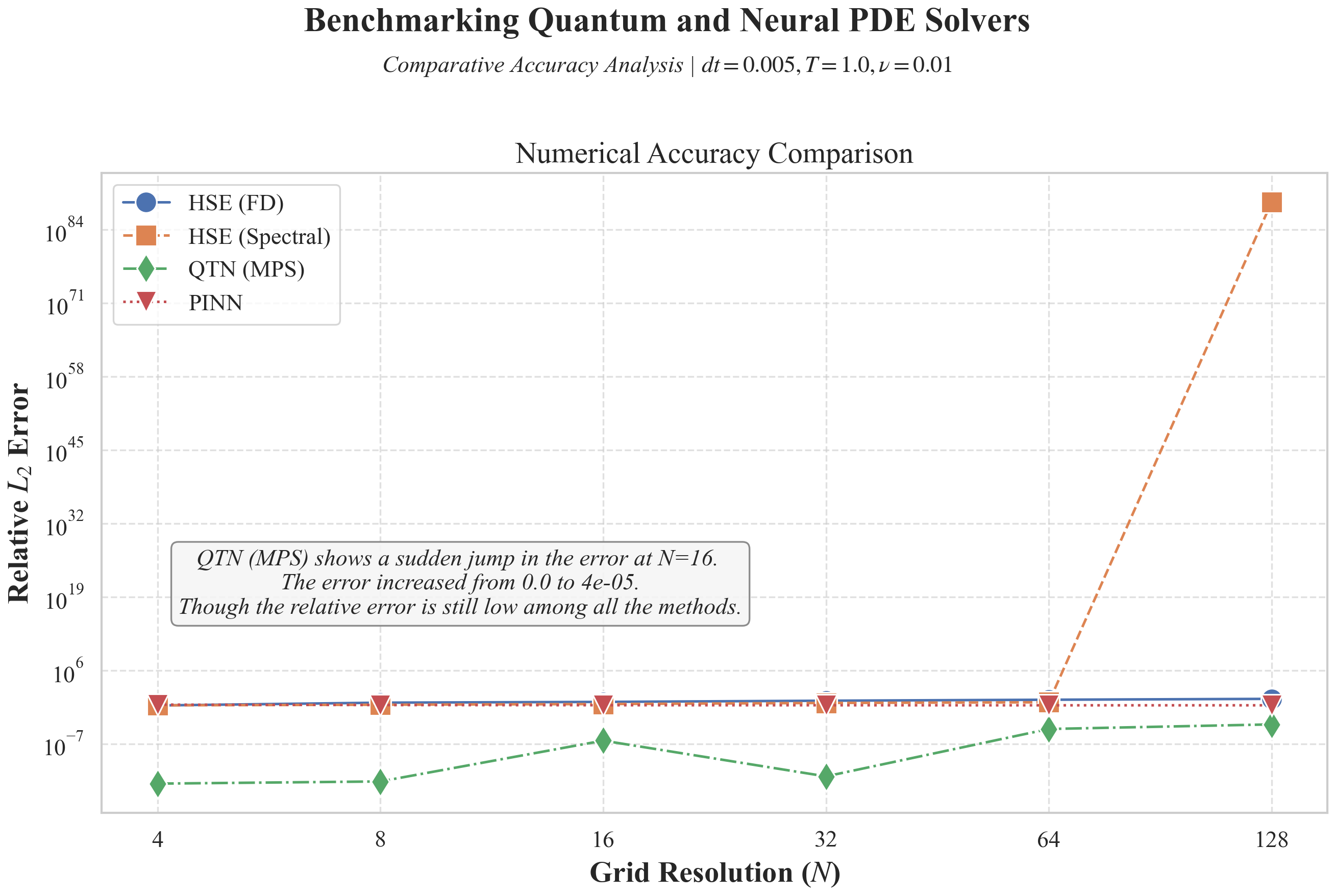}
    \caption{Figure shows the relative $L_2$ error for grid sizes from $N=4$ to $N=128$}
    \label{fig:accuracy}
\end{figure}

The table \ref{runtime_complexity} gives the runtime complexity and memory utilisation of both classical and quantum algorithms. HSE has polylogarithmic complexity in runtime, key potential for quantum advantage. Though in the case of HSE-spectral, dense Hamiltonian limits this scaling. 
Meanwhile, the QTN runtime scales logarithmically with the system size $N$ but is heavily dependent on the bond dimension $\chi$, following an $\mathcal{O}(n \cdot \chi^4)$ relation.

 \subsection{Numerical Accuracy \& Convergence}
 Figure \ref{fig:accuracy} shows the variation of the $L_2$ error with grid resolution $N$. The simulation is run from $t=0$ to $t=0.1$ with a step size of $dt=0.005$. QTN achieve a lower error ($L_2 \sim 10^{-7}$) and near-constant runtime scaling, effectively capturing shock fronts. While HSE (FD) and PINN maintain stability and capture the shock, the spectral method diverges significantly.
\begin{figure}[ht!]
    \centering
    \includegraphics[width=1\columnwidth]{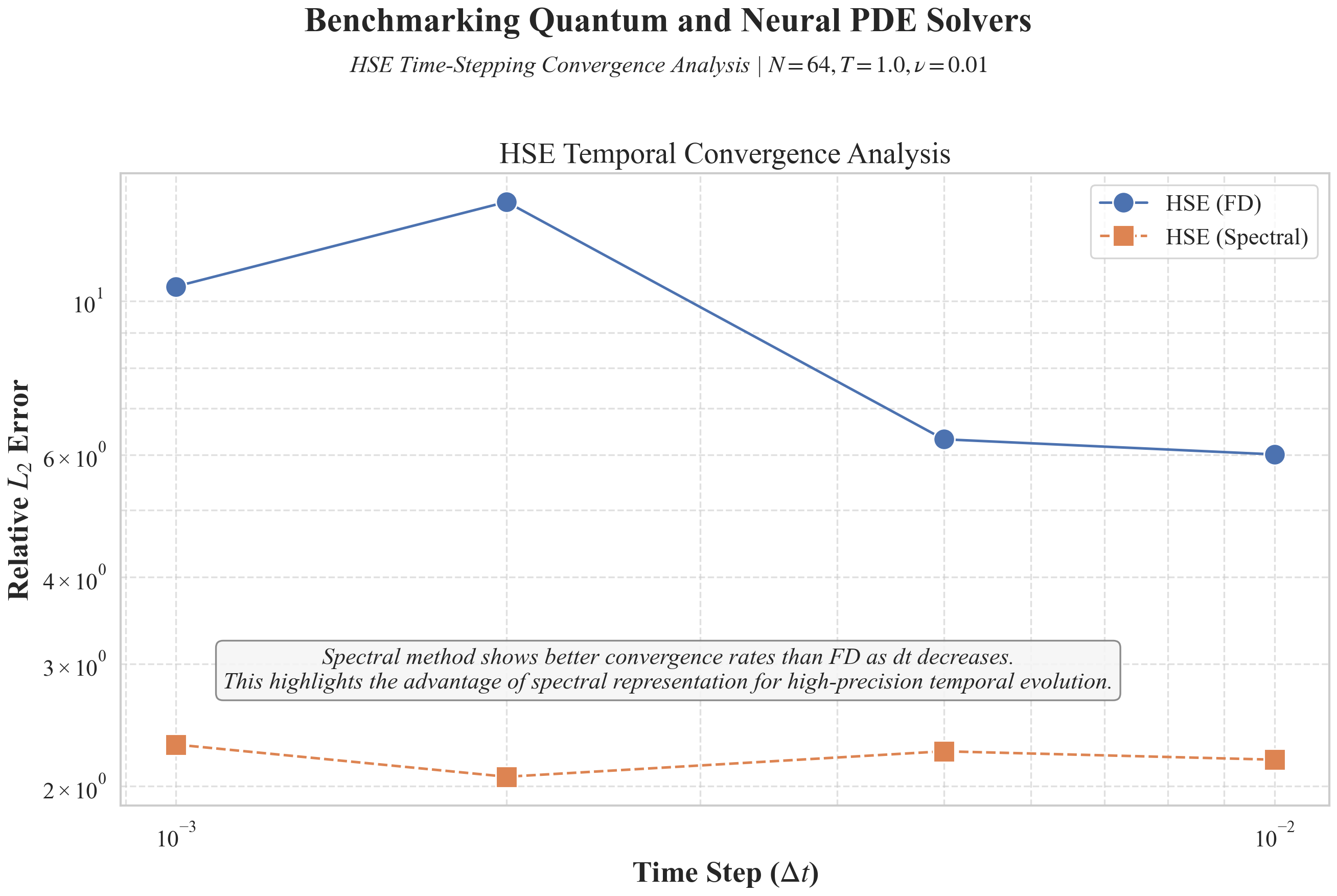}
    \caption{HSE L2 Error convergence with Trotter step size $dt$.}
    \label{fig:hse_l2_with_dt}
\end{figure}

The accuracy of the HSE solver is primarily constrained by artefacts arising from the Trotter decomposition. As illustrated in Figure \ref{fig:hse_l2_with_dt}, the $L_2$ error converges as the time step $dt$ is reduced. Although the algorithm offers a direct mapping of the fluid variables to a quantum wavefunction, the computational overhead associated with state initialisation and the $\mathcal{O}(N)$ measurements required for readout remains a significant bottleneck.
\begin{figure}[ht!]
    \centering
    \includegraphics[width=1\columnwidth]{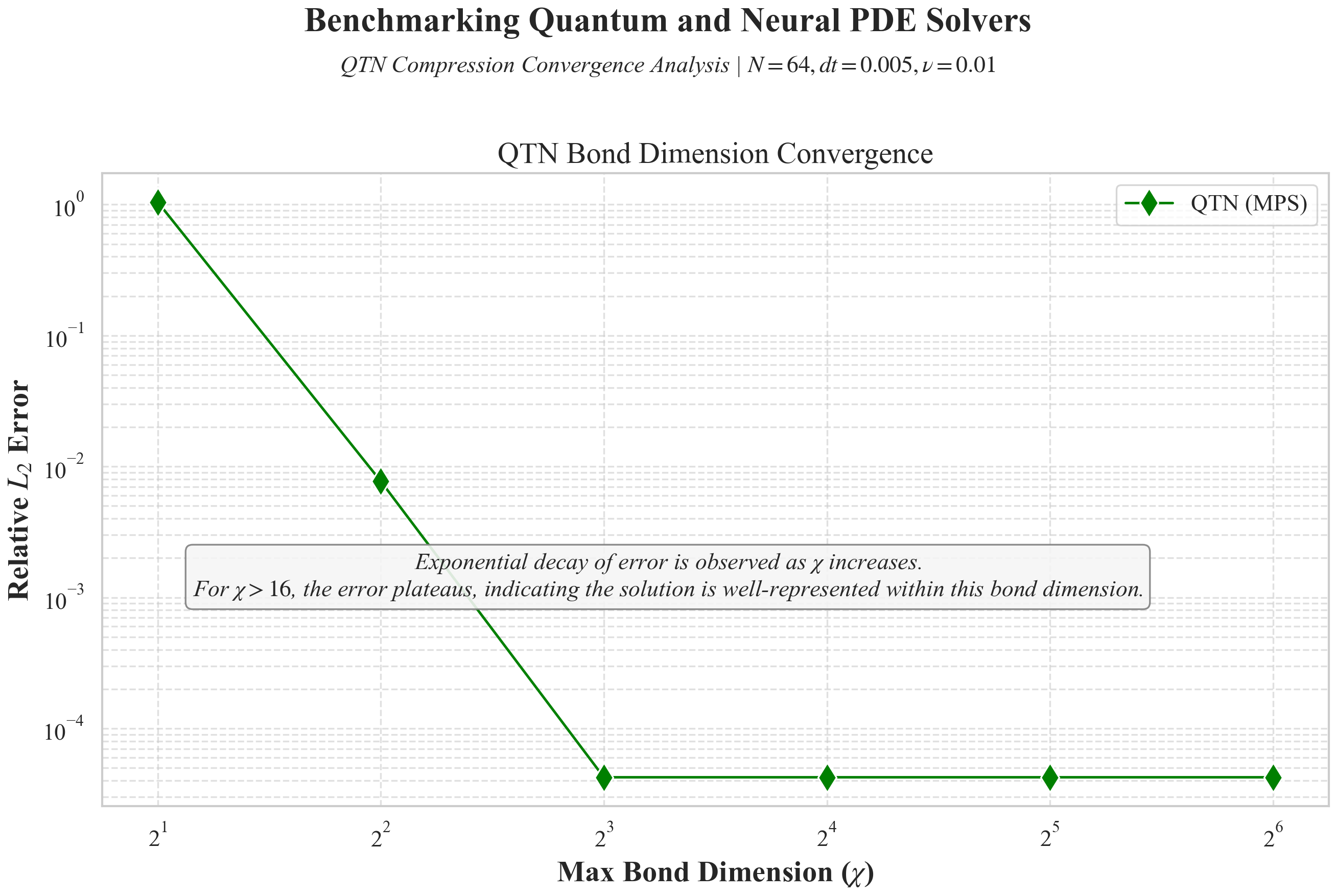}
    \caption{QTN Accuracy vs Bond Dimension $\chi$.}
    \label{fig:accuracy_chi}
\end{figure}

\begin{figure}[ht!]
    \centering
    \includegraphics[width=\columnwidth]{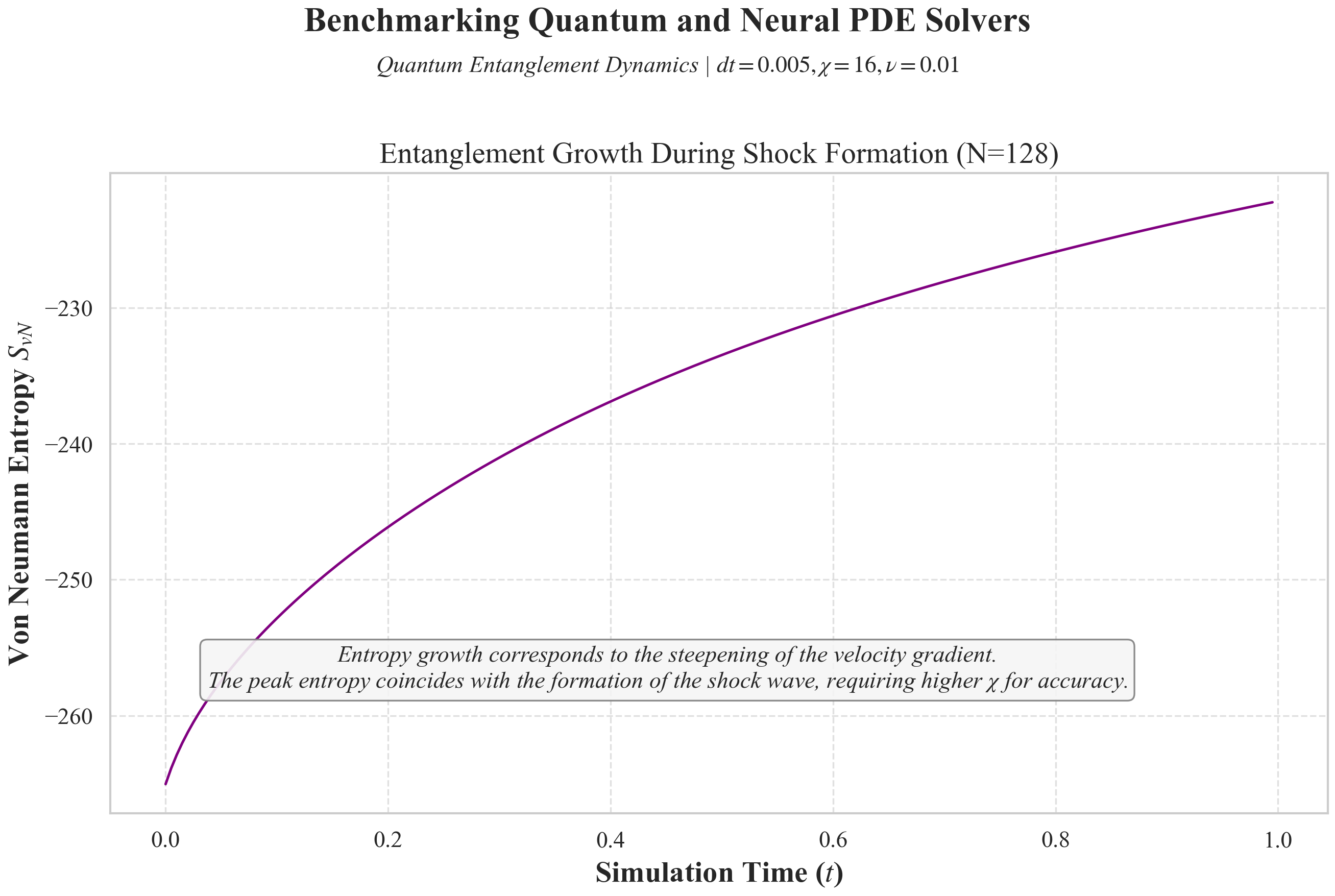}
    \caption{Growth of entanglement entropy over time in the QTN solver as the shock forms.}
    \label{fig:entanglement}
\end{figure}

In contrast, the performance of the QTN solver is dictated by the bond dimension $\chi$, as depicted in Figure \ref{fig:accuracy_chi}. Expanding $\chi$ enables the tensor network to accommodate larger entanglement scaling, achieving error norms comparable to classical Finite Difference methods ($\approx 10^{-4}$) when $\chi \ge 16$. However, this compression scheme faces challenges during shock formation, which induces a rapid growth in entanglement entropy (refer to Figure \ref{fig:entanglement}).

\subsection{Resource Scaling}
\begin{figure}[ht!]
    \centering
    \includegraphics[width=1\columnwidth]{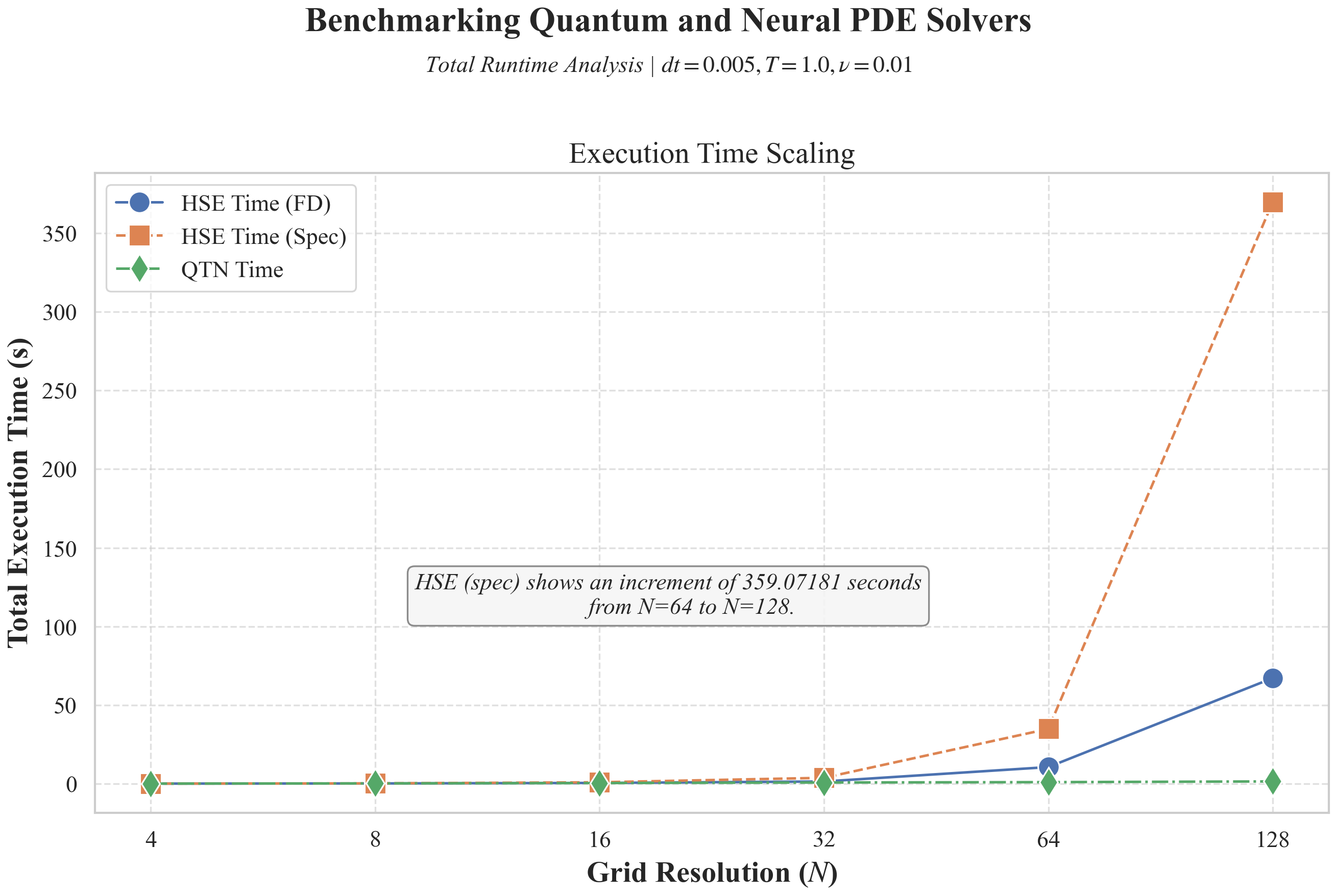}
    \caption{Figure shows the total time of execution of HSE and QTN algorithm with grid resolution $N=4$ to $N=128$}
    \label{fig:timeevolution}
\end{figure}

Figure \ref{fig:timeevolution} benchmarks the execution runtime. QTN bypasses in runtime efficiency that severely hamper the spectral HSE approach for $N>32$. For $N>64$, handling dense spectral operators on classical simulators becomes more expensive than the sparse difference operators.

\begin{figure}[h!]
    \centering
    \includegraphics[width=1\columnwidth]{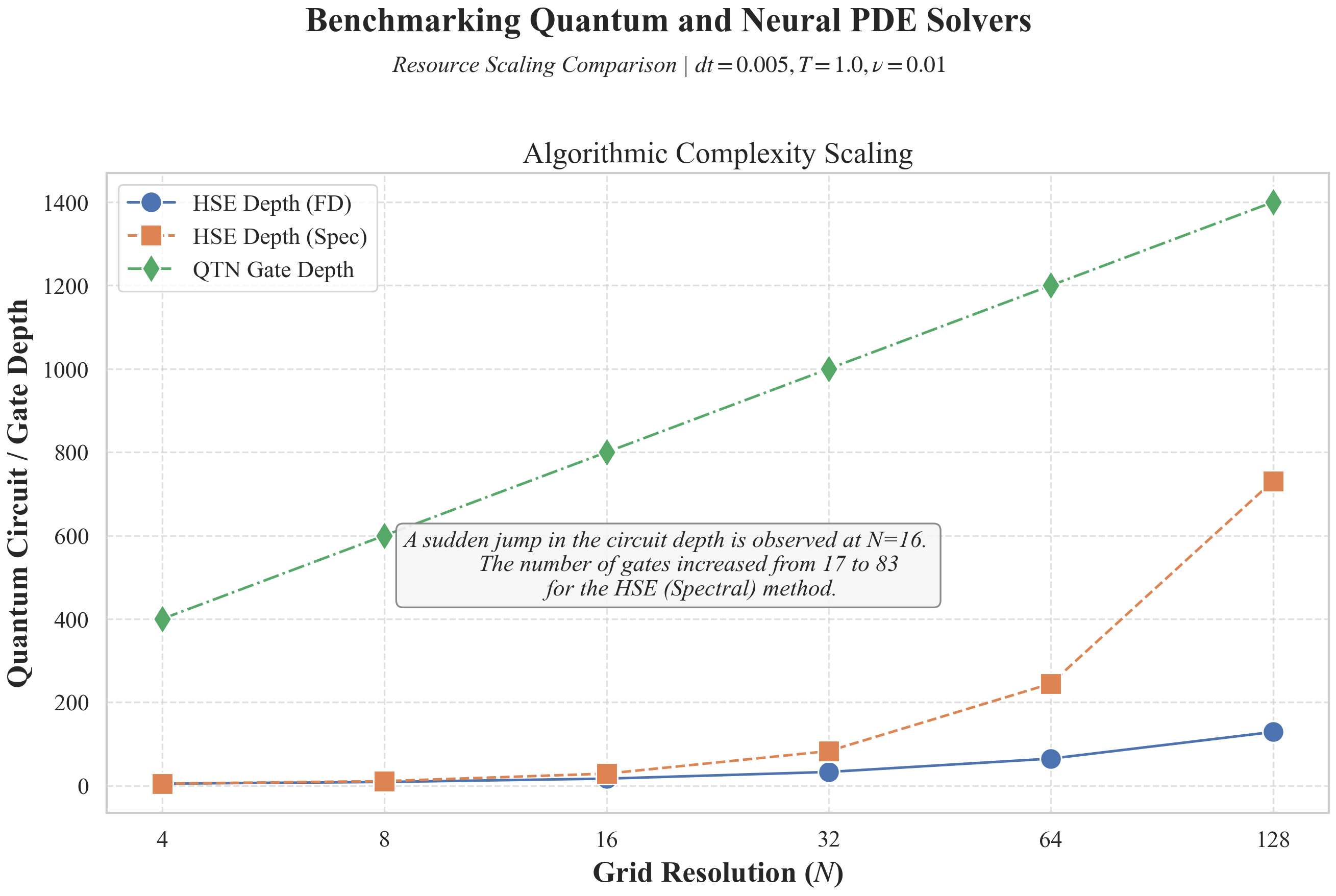}
    \caption{Figure shows the variation of the circuit depth with grid resolution $N=4$ to $N=128$}
    \label{fig:circuitdepth}
\end{figure}

The circuit depth variation for HSE and QTN is shown in Figure \ref{fig:circuitdepth}. For HSE(spectral), the depth for $N>64$, which correlated to previous observations of all-to-all Hamiltonians are ill-suited for NISQ devices. The FD approach shows manageable polynomial scaling, maintaining a low depth $<150$ gates at $N=128$. The QTN shows a linear relation, $(D \propto N$), reaching $\approx 1400$ gates at $N=128$. We assumed an \textit{ad-hoc} linear assumption to allow the system to effectively capture volume-law entanglement generated in non-equilibrium shock ranges, prioritising robustness over maximum compression.

\begin{figure}[ht!]
    \centering
    \includegraphics[width=\columnwidth]{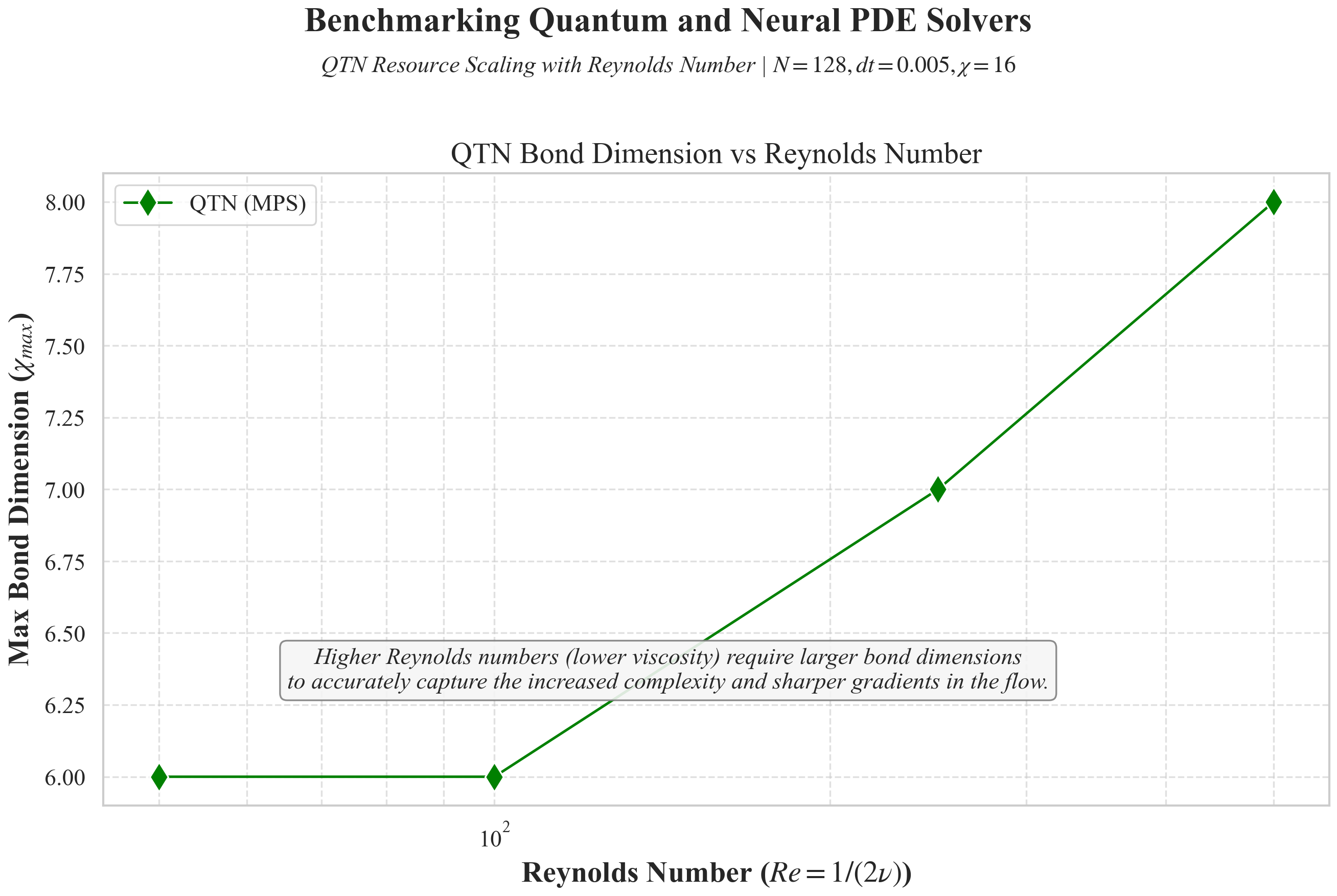}
    \caption{QTN Bond Dimension vs Reynolds Number}
    \label{fig:qtn_bond_dim_vs_re}
\end{figure}

\subsection{Reynolds Scaling:}
Figure \ref{fig:qtn_bond_dim_vs_re} shows the maximum bond dimension $\chi_{max}$ required to maintain accuracy as the Reynolds number increases for $N=128$. Our result recovers the expected physical scaling, with $\chi$ increasing from 6.0 to 8.0 as $Re$ rises. This aligns with the work by Lubasch et al. (2020) and Gourianov et al. (2022), establishing that for converged simulations, the bond dimension must scale logarithmically or polynomially with inverse viscosity ($\chi \sim \log(Re)$) to capture the fine-scale correlations of the thinning shock front. For a coarse grid, as in our case, the bond dimension remains constant. This behaviour is consistent with the properties of MPS in the coarse-grid regime. 

\begin{figure}[ht!]
    \centering
    \includegraphics[width=\columnwidth]{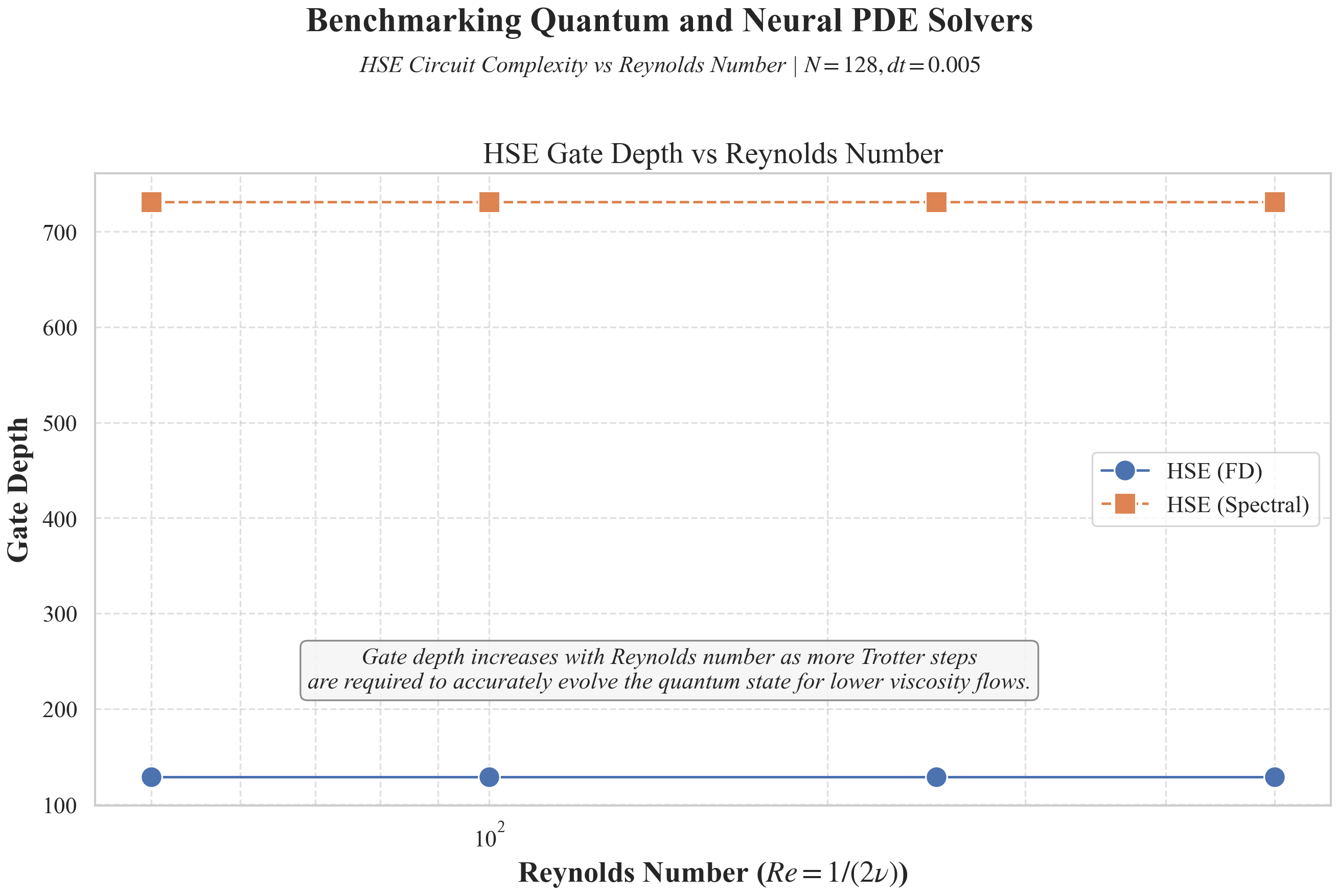}
    \caption{HSE Circuit Complexity vs Reynolds Number}
    \label{fig:hse_gate_depth_vs_re}
\end{figure}

Figure \ref{fig:hse_gate_depth_vs_re} shows a constant gate-depth for both Finite Difference and Spectral implementations for different $Re$. The theoretical studies by Child et al. \cite{childs2021theory} often predict that the total query complexity scales with $Re$. Our results show the gate depth per step remains a deterministic, hardware-friendly constant.

\section{Conclusion}
\label{sec:conclusion}
We have presented a diagnostic benchmark of quantum-inspired (QTN), quantum-native (HSE) and deep learning (PINN) solvers for 1D Burger's equation. Our findings indicate that: 
\begin{itemize}
    \item \textbf{Quantum Tensor Networks (QTN)} emerged as the optimal strategy for the current NISQ era, yet their advantage is regime-dependent. Our results indicate that QTN offers exponential compression primarily in low-entanglement, laminar regimes. In shock-dominated flows, the method encounters an "entanglement barrier" where the formation of sharp discontinuities necessitates a rapid growth in bond dimension to resolve fine-scale viscous structures. Consequently, while QTNs remain efficient for smooth flows, the resource requirements for resolving shocks drive the complexity back towards classical scaling limits, highlighting a critical boundary for quantum advantage in non-linear dynamics.
    \item \textbf{Hydrodynamic Schrödinger Equation (HSE) }provides a conceptually elegant framework for directly mapping fluid variables to quantum hardware. However, it is practically constrained by the "input/output bottleneck," specifically the readout costs scaling as $\mathcal{O}(N)$ required to update the non-linear potential at every time step. Furthermore, while spectral discretisations offer theoretical high-order accuracy, we demonstrated that they are exponentially expensive on gate-based quantum computers due to the dense connectivity required, resulting in prohibitive circuit depths.
    \item \textbf{Physics Informed Neural Networks} offers a flexible, mesh-free alternative but currently lacks the precision of grid-based solvers for high-Reynolds number flows due to optimisation hardness. 
    
\end{itemize}
This work shows that classical algorithms still offer the best speed and precision in this regime. While Quantum-inspired and Quantum-native show interesting representational advantages, they do not currently yield a computational advantage for classical fluid dynamics. On the other side, Physics-Informed Neural-Network offers a mesh-free flexibility but requires spatio-temporal collocation sampling, creating a training bottleneck similar to mesh refinement. 

\section{Future Works}
While this study establishes a robust baseline for quantum fluid simulation using the 1D Burgers' equation, several avenues remain for extending the practical utility of these algorithms.
\begin{itemize}
    \item \textbf{Hardware Deployment:} The current benchmarking relied on state-vector and noisy density-matrix simulators. A critical next step is to deploy the QTN and HSE architectures on physical Quantum Processing Units (QPUs), such as superconducting or trapped-ion processors. Real hardware execution is essential to validate the effective gate depths and to quantify the impact of physical connectivity constraints and decoherence on solution fidelity.
    
    \item \textbf{Extension to Higher Dimensions:} Extending to 2D and 3D flows is necessary to capture complex fluid phenomena such as vortex shedding and turbulence. Future work will focus on extending the tensor network ansatz to 2D geometries to model the multi-dimensional Navier-Stokes equations. 
    
    \item \textbf{Long-Time Integration and Stability:} Our current simulations were limited to short temporal windows ($t \in [0, 1s]$). To study steady-state behaviours and fully developed turbulence, the temporal domain must be significantly extended. 
    
    \item \textbf{Real-World Use Cases:} Finally, we aim to push the simulation limits towards industrially relevant regimes. This involves stress-testing the algorithms at significantly higher Reynolds numbers to understand the "quantum advantage" threshold for turbulent flows. This is particularly important for industries including aviation, defence, space and simulation.

    \item \textbf{System Resource Profiling}: Our current simulations were performed on high-performance classical architectures (Apple M4 / Intel Core i9 CPU + GPU hybrid systems). However, we have not yet conducted a deep dive into the granular system resource usage, specifically energy consumption, peak memory bandwidth, and thermal overheads. Future studies should quantify the running costs to better define the crossover point where a physical QPU becomes energetically favourable over a GPU-accelerated emulation.
    
\end{itemize}

\section{Code And Data Availability}
The source code and trained models used in the study are available upon request from the corresponding authors.

\section{Acknowledgement}
The authors gratefully acknowledge The Washington Institute for STEM Entrepreneurship and Research for the opportunity to participate in the Quantum Algorithms for Differential Equations, 2025. We specifically thank the organisers for providing the initial problem statement, which served as the foundational inspiration for this comparative study.

\section{AI Usage}
During the preparation of this work, the authors used large language models (Google Gemini web platform, \href{https://gemini.google.com/app}{https://gemini.google.com/app}) to assist with grammatical editing, linguistic refinement, and LaTeX formatting of the manuscript. The authors reviewed and edited the content as needed and take full responsibility for the content of the publication. No AI tools were used to generate the underlying scientific data or the primary results of the benchmarks presented herein.

\bibliographystyle{IEEEtran}
\bibliography{bibtex} 

\end{document}